\documentstyle[aps,preprint]{revtex}

\begin{document}

\draft

\title{Evidence of magnetization-dependent polaron formation
in La$_{1-x}A_x$MnO$_3$, $A$=Ca, Pb
}

\author{C. H. Booth$^1$, F. Bridges$^1$, G. J. Snyder$^2$, T. Geballe$^2$
}

\vspace{0.5ex}

\address{$^1$Department of Physics, University of California,
Santa Cruz, CA 95064}

\address{$^2$Department of Applied Physics, Stanford University, Stanford,
CA 94305}

\date{draft: \today}

\maketitle

\begin{abstract}
X-ray-absorption fine-structure measurements at the Mn $K$-edge as a function 
of temperature were performed on samples of 
La$_{1-x}$Ca$_x$MnO$_3$ and La$_{0.67}$Pb$_{0.33}$MnO$_3$.  
All samples have a metal-insulator (MI) transition near the ferromagnetic 
transition, except the Ca$_{0.5}$ sample, which does not have a MI transition.
Near $T_c$ for the MI samples, the Debye-Waller $\sigma^2$ for 
the Mn-O atom-pair increases rapidly with temperature.  This non-thermal 
disorder is mostly along the
direction of the neighboring oxygens.
These results strongly suggest small polarons which are delocalized by a
finite magnetization.
\end{abstract}





Although the basic physics behind the connection between the ferromagnetic (FM)
transition, the conduction and the ``colossal''
magnetoresistance (CMR) in materials such as La$_{1-x}A_x$MnO$_3$ ($A$= one 
of certain divalent metals) is described by the double-exchange mechanism of 
Zener\cite{Zener51}, several features of the CMR materials are
inconsistent with 
a model that only consists of double exchange\cite{Millis95}.  Including
spin-polarons and electron-electron interactions can help explain some
features, such as the absolute value of $T_c$\cite{Varma96}.  
Including thermal spin fluctuations may account for the 
conductivity\cite{Furukawa95a}.  However, as pointed out by 
deGennes\cite{deGennes60}, calculations of the conductivity may
be inaccurate unless they include lattice interactions.
One 
possibility is that the charge-hopping in the double-exchange model induces
polaron formation\cite{Millis95,Millis96}.  Such a formation would invalidate 
the assumption that the mean-free
path is independent of the carrier mass, thereby enhancing the 
spin-spin coupling term in the double-exchange Hamiltonian\cite{Millis95}.
If a polaron mechanism is to explain the discrepancies between 
the CMR effect with the double-exchange model, polarons must exist at least 
above
$T_c$, and undergo a change in their dynamics below $T_c$ with the result that
the root-mean-square (rms) displacements of the atoms involved must be 
significantly reduced.  Alternatively, if such changes are not observed,
thermal spin fluctuations may be a better description.

Above $T_c$, the
primary form of charge-transport will be hopping, while below $T_c$
the system will be in a metallic phase.  In the former case, a moving charge
may carry a lattice distortion along with it (small polaron), while in 
the latter case the lattice will not have time to react or may distort on a
larger scale (large polaron).


Experimental evidence for polaron distortions in these materials is 
mounting\cite{Kwei96,Billinge96,Louca96}.
Neutron diffraction measurements on La$_{1-x}$Ca$_x$MnO$_3$ ($x$=0.12, 0.21,
and 0.25) have shown Jahn-Teller distortions of the O(2) (planar) atoms, 
producing three Mn-O bond lengths in the planes separated by about 0.02 \AA.
Both the planar and axial (O(1)) oxygens have changes in the slope of the
rms displacements with temperature at $T_c$ in samples with metal-insulator 
(MI) transitions 
for the O(1), O(2)  and La/Ca sites\cite{Kwei96}.  

A more
compelling measurement should be provided by an experiment than can act as a 
{\it local} probe, since measurements of magnetization do not depend on 
long-range order.  Such probes give information about the correlations between 
atomic positions\cite{Billinge96,Booth95}.  A pair
distribution function analysis (PDF) of the same data as the diffraction
analysis described above shows an increase of 0.12 \AA{ } 
in the distribution width of the Mn-O and O-O pairs\cite{Billinge96}.  Interestingly,
the width in temperature of the transition is much broader in the PDF
analysis, suggesting that the effect is present on a local scale far
away from $T_c$.  Another
neutron diffraction study claims to see an anomalously long Mn-O peak at 
$\sim$2.28 \AA, which shows no temperature dependence, on a sample of 
La$_{0.67}$Sr$_{0.33}$MnO$_3$.  
Previous x-ray-absorption fine-structure (XAFS)
measurements have been somewhat inconsistent
with these studies, measuring a change in the shape of the Mn-O distribution
across $T_c$\cite{Tyson96}, and a long Mn-O bond of 
$\sim$2.5 \AA\cite{Tyson96b}.

This work reports our XAFS results for the Mn-O atom pair.
The data show that the smooth and rapid change in the Debye-Waller
broadening parameter $\sigma^2$
near $T_c$ for this bond as seen in PDF experiments\cite{Billinge96} on 
$A$=Ca$_{0.21}$ and Ca$_{0.25}$, is  
present in our XAFS experiments on $A$=Ca$_{0.25}$, Ca$_{0.33}$ and Pb$_{0.33}$.
This change is shown to be inconsistent with broadening due to thermal phonons.
We also measure $\sigma^2$ for the next two near-neighbors in the Ca$_{0.25}$ sample, 
and find the changes near $T_c$ are smaller in the Mn-Ca/La 
scattering paths than in the Mn-O and the Mn-O-Mn paths,
and therefore appear to be mostly along the crystal axes for these 
materials.  Measurements of a sample which also has an antiferromagnetic 
transition but no metal-insulator transition ($A$=Ca$_{0.5}$) show
no such effects, consistent with the insulating Ca$_{0.12}$ sample in the 
neutron scattering study.


Powder samples were made by the solid-state reaction of various proportions of 
the elemental oxides 
MnO, La$_2$O$_3$, and PbO and the carbonate CaCO$_3$. Further details can 
be found in Ref. \cite{Snyder96}.
Magnetization vs. $T$ for the samples are shown in
Fig. \ref{mag_fig}.  $T_c$'s (estimated from 1/2 saturation
magnetization) are 236$\pm$5 K,
268$\pm$5 K, and 241$\pm$5 K for the $A$=Ca$_{0.25}$, Ca$_{0.33}$, and 
Pb$_{0.33}$ samples, respectively.
The Ca$_{0.5}$ sample's $T_c$ is 240$\pm$10 K; it also has a
N\'{e}el temperature around 210 K.

All absorption measurements were made in the transmission mode
at the Stanford Synchrotron Radiation
Laboratory on beamline (BL) 2-3 with Si(220) monochomator crystals and BL 10-2 
with Si(111) crystals.  Samples were ground, run through a 40 $\mu$m
sieve and brushed onto scotch tape.  Four layers of such tape were stacked to
provide samples with a Mn $K$-edge step of roughly 0.3.
Samples were then placed in an Oxford helium-flow cryostat.  Temperature
was monitored via a sensor on the probe about 2-6 cm from the sample.  
The temperature was regulated within 0.1 K, but the sample temperature could be
as much as 2 K higher than the nominal temperature, especially for the
higher temperatures.  

Data were reduced following standard 
procedures\cite{Hayes82,Li95b,Bridges95b}.  The XAFS were isolated 
by defining the parameter $\chi(k)=\mu(k)/\mu_0(k) - 1$,
where $k$ is the magnitude of the photoelectron wave vector, $\mu(k)$
is the total x-ray absorption due to the Mn $K$-shell excitation, and $\mu_0(k)$
is the part of $\mu(k)$ that does not include the interference of the outgoing
part of the photoelectron wave function and the backscattered part.
Although the monochromator crystals were more than 50\% detuned, measurements 
of $\chi(k)$ from BL 2-3 were slightly lower in 
amplitude ($\sim$3\%) than the measurements from BL 10-2 at the same 
temperature, presumably from second-harmonic contamination.
We therefore multiplied the data from BL 2-3 by 1.03 to allow better comparisons
of the two data sets.  

Fits to the data utilized theoretical standards as calculated by 
FEFF6\cite{FEFF6}.  Theoretical standards (as opposed to standards
obtained experimentally) allow for the measurement of an absolute broadening
factor in the pair-distribution function, and thus allow for a direct comparison
to diffraction results.  


Fig. \ref{rs_fig} shows Fourier transforms (FT) of $k\chi(k)$ for each sample
measured at 100 K and at 300 K.  The first peak in these transforms is due to
the Mn-O atom pair, while the broad peak centered near 3.2 \AA{ } is really
a multipeak with contributions from Mn-$A$, Mn-La and the Mn-Mn pairs.  The
main feature to recognize in this figure is that the MI 
materials, all show a reduction in the Mn-O peak 
as $T$ is raised from 100 K to 
300 K.  Such behavior may indicate a {\it soft} bond with a low
Debye temperature, or possibly some other kind of change in the structural 
parameters.  On the other hand, the Mn-O peak in the 
La$_{0.5}$Ca$_{0.5}$MnO$_3$ data is nearly unchanged (even near the 
FM transition), suggesting a relatively high Debye temperature and no other 
structural changes.

We fit each spectrum to standards of Mn-O, Mn-La, Mn-$A$, and the multiple 
scattering
Mn-O-Mn path.  These paths were calculated assuming the simple perovskiite
structure with no distortions.  In particular, the Mn-O-Mn bond angle was
taken to be 180$^\circ$, even though for some compounds in the LaMnO$_3$
series, this angle can be as small as 160$^\circ$.  This angle is expected to
be close to 180$^\circ$ for these materials\cite{Mahendiran96}.
The amplitude reduction factor $S_0^2$ (an overall coefficient not accounted 
for by FEFF6) was based on the Mn-O peak for
each sample and varied between 0.70 and 0.75, assuming 6 oxygens in the Mn-O
peak.  The Mn-O-Mn path was allowed to have its own $S_0^2$ of 0.55 to account
for any possible change in the bond angle and for problems with FEFF6's 
multiple scattering calculations.

Fits to the Mn-O peak for the lowest temperature data are very similar for 
all samples.  None of the bond lengths changed with temperature
within the estimated error.  The Mn-O bonds for the $A$=Ca$_{0.25}$, Ca$_{0.33}$,
Ca$_{0.5}$, and Pb$_{0.33}$ are 1.95$\pm$0.005 \AA, 1.95$\pm$0.005 \AA, 
1.92$\pm$0.005 \AA{ } and 1.96$\pm$0.005 \AA, respectively.  These results are 
consistent with diffraction studies\cite{Kwei96,Mahendiran96,Radaelli95}.
Fits to the Mn-O broadening parameter for these data are presented in 
Fig. \ref{s2_all_fig} and are consistent with the raw data in Fig. \ref{rs_fig},
as well as clearly showing a change in the local Mn-O 
environment.
The $\sigma^2$ from the La$_{0.5}$Ca$_{0.5}$MnO$_3$ data is fit reasonably 
to a correlated-Debye model plus static disorder with 
$\Theta_D \cong$ 950 K.
Interestingly, none of the data from the MI materials fits a Debye model 
very well in any temperature range.  Each
one starts out at the lowest temperatures with a $\sigma^2$ lower than
the 50\% Ca material (note that the Ca$_{0.5}$ results are shifted in 
Fig. \ref{s2_all_fig}), suggesting a higher Debye temperature or less
static disorder.  However,
as $T$ is increased (but still below the Curie point) $\sigma^2$ increases
too sharply to be consistent with such a high $\Theta_D$.  A lower $\Theta_D$
would be possible except that an unphysically negative amount of static
disorder would be necessary to describe the data.  Within
$\sim$40 K of $T_c$, $\sigma^2$ begins to climb even more sharply, 
leveling off about 40 K above $T_c$.  This behavior is fundamentally 
different from the behavior expected for phonon broadening with either 
a Debye or Einstein model, including static disorder.  For these models, 
the tangent line to the data always has a positive intercept.  A straight line 
though the data in the transition region has a negative intercept. 


Fits to the further neighbor Mn-La, Mn-Ca and Mn-O-Mn paths for the
La$_{0.75}$Ca$_{0.25}$MnO$_3$ data are presented
in Fig. \ref{ca25_fig}.  The relative intensity of the Mn-La and Mn-Ca
signal could vary widely in the fits; interferences occur which
make various combinations possible.  Therefore, their intensities were held
at a fixed 3:1 ratio, and their bond lengths were held equal in the fits.
This latter constraint is probably a good assumption, given the behavior of
Ga$_{1-x}$In$_{x}$As\cite{Mikkelsen83}.  The Mn-Ca/La bond length was 
measured to be 3.38$\pm$0.02 \AA{ } and did not change with temperature.  
All bonds in Fig. \ref{ca25_fig} show similar increases of $\sigma^2$
near $T_c$.  We obtain a rough measure of the size of the increase in
$\sigma^2$ by extrapolating the low temperature fits to $\sigma^2$ past 
$T_c$ and measuring the maximum difference between the data and this 
extrapolation.  For the Mn-O bond in the 25\% Ca sample, this increase is
approximately 0.0022 \AA$^2$, corresponding to an increase in
$\sigma$ of 0.047 \AA{ } through $T_c$.  (Note that $\sigma$'s add in
quadrature.) 
If the disorder in the Mn-O pair is only along the $a$, $b$ and $c$
axes, its projection onto the Mn-Ca/La pair is 
$\sim$ $\Delta \sigma^2 \times \frac{r_{Mn-O}}{r_{Mn-La/Ca}}$
$\sim$ 0.0022 $\times (\frac{1.93}{3.37})^2$=0.0007 \AA$^2$ (ignoring any oxygen
dimpling).  The measured
size of the anomaly for the Mn-La/Ca pair is, in fact, 0.0007 \AA$^2$.
This analysis suggests that either (a) only the Mn is distorting (unlikely,
given that Mn is nearly 4 times heavier than oxygen, and there is no step in 
the Mn thermal parameters in Ref. \cite{Kwei96}) or (b) the Ca does 
distort, but only a fraction of the amount that the oxygens distort.  In 
either case, we conclude that the predominant direction of the increased 
disorder is along the crystal axes.  This conclusion is supported by the 
0.0028 \AA$^2$ jump for the Mn-O-Mn path.  This path may have a larger step in 
$\sigma^2$ due to either a change in the correlation of the Mn positions or
deviations of the bond angle from 180$^\circ$; these data cannot differentiate 
between these two possibilities.  La-edge data should help further determine
the nature of this disorder.


It is interesting to note the similarities between the magnetization data and
the $\sigma^2$ fit results.  Each distribution shows a long tail towards low
temperature and approximately the same width in the transition region.  
Considering the magnetic data together with the high ($\sim$ 1000 K) Debye 
temperature necessary to describe the XAFS at the lowest temperatures, we are 
lead to 
the conclusion that virtually {\it all} the temperature dependence demonstrated
by the MI materials is magnetic in origin and caused by the degree
of localization of polarons.  At low temperatures and high magnetizations,
the conduction electrons are allowed to hop freely between spin-aligned ions,
and thus the polarons are effectively delocalized. Contrary to the usual
thermal activation of polaron hopping, as $T$ is increased towards $T_c$ and the
magnetization decreases, the polarons encounter more and more resistance
in the form of unaligned ion-spins and are thus more likely to become trapped,
causing more localization and distortion.  This trend will eventually
reverse as $T$ increases.

Qualitatively, this distortion is consistent with Millis' prediction that
the polaron distortion should cause an
increase in the width of the distribution of the oxygen atoms near $T_c$
of order 0.1 \AA\cite{Millis95}.  

These data and this analysis are not consistent with a previous XAFS
study\cite{Tyson96}.  In that study, Tyson {\it et al} measured a drastic
change in the shape of the XAFS spectrum between $T$=80 K and 273 K, which
they interpreted
as a change in the Mn-O bond length distribution.  As is clear from
Fig. \ref{rs_fig}, we do not measure
a fundamental difference between $T$=100 and 300 K (or anywhere in between),
although there is an obvious difference in the amount of disorder.  In another
study, Tyson {\it et al} measure a long Mn-O pair at 
$\sim$2.5 \AA\cite{Tyson96b}, which
is roughly in agreement with Louca {\it et al}\cite{Louca96}.  We do see some
evidence for such a peak, however there are several other interpretations,
including an analysis artifact arising from FEFF6, and therefore
do not include it in this analysis.  Including this extra peak does not
affect the results of this paper.

In conclusion, this work demonstrates a clear step-like increase in the width 
of the Mn-O bond near $T_c$ for samples with a MI transition and a 
temperature dependence below $T_c$ that is related to the magnetization of the 
sample.  Since diffraction measurements do not show any step in the
width parameters of any sites (they do show a change in slope at $T_c$)\cite{Kwei96},
the steps we have measured in this work can be interpreted as arising from
a more negatively-correlated character in the positions of the atoms in
the Mn-O and
Mn-Mn pairs.  In addition, we have shown that most of these displacements
are along the crystal axes, as one expects from a polaron distortion.
Lastly, we should point out that since diffraction measures 
essentially the 
same distortions above and below $T_c$, consistent with Jahn-Teller 
band-splitting, the measurements reported in this work are separate and 
distinct from a change in the Jahn-Teller distortions.

We wish to thank G. Kwei for useful discussions and assistance in collecting
the data.
The experiments were 
performed at the Stanford Synchrotron Radiation
Laboratory, which is operated by the U.S. Department of Energy, Division
of Chemical Sciences, and by the NIH, Biomedical Resource Technology Program,
Division of Research Resources.  The experiment was partially carried out on
UC/National Laboratories PRT beam time. 
The work is supported in part by NSF grant DMR-92-05204.



\begin{figure}
\caption{Magnetization vs. $T$ for all samples. The applied fields were
(from top to bottom) 2.5 kOe, 10 kOe, 2.5 kOe and 5 kOe.  The magnetization data
for the Ca$_{0.25}$ and Ca$_{0.50}$ curves have been multiplied by 4 and 15,
respectively.}
\label{mag_fig}
\end{figure}

\begin{figure}
\caption{$r$-space data of each sample at (solid) $T$=100 K and (dotted) 
$T$=300 K.  Data are transformed from 3.5-14.5 \AA$^{-1}$ and Gaussian broadening
by 0.3 \AA$^{-1}$.}
\label{rs_fig}
\end{figure}

\begin{figure}
\caption{$\sigma^2$ vs. T for the Mn-O peak for all samples measured.
The Ca$_{0.5}$ measurements are shifted down by 0.001 \AA$^2$ for clarity.
Error bars indicate reproducibility of the data and fit, and in no way
attempt to estimate systematic errors, with may be as large as 20\%.  The dotted line
shows a correlated-Debye model for $\Theta_D$=950 K with 0.0011 \AA${^2}$ 
static broadening, also shifted down by 0.001 \AA$^2$.}
\label{s2_all_fig}
\end{figure}

\begin{figure}
\caption{$\sigma^2$ vs. $T$ for the La$_{0.75}$Ca$_{0.25}$MnO$_3$ Mn-O, 
Mn-La/Ca and Mn-O-Mn scattering paths.  The vertical dotted line marks $T_c$.
}
\label{ca25_fig}
\end{figure}

\end{document}